# Students know AI should not replace thinking, but how do they regulate it? The TACO Framework for Human–AI Cognitive Partnership

Cecilia Ka Yuk Chan

Affiliation: The University of Hong Kong

Email: ckchan09@hku.hk

Website: https://tlerg.talic.hku.hk/

https://aiedlab.hku.hk/

**Abstract**

As generative artificial intelligence becomes increasingly embedded in educational practice, a central concern is whether students use AI as cognitive support or as a substitute for thinking. Prior research shows that learners recognise this boundary conceptually and acknowledge that "AI should not replace thinking." However, whether such awareness translates into structured regulation during actual AI use remains unclear.

Drawing on data from Hong Kong secondary students, this study examines how learners perceive their management of the boundary between assistance and outsourcing in practice. Findings show that awareness did not consistently translate into regulation; ethical belief did not necessarily lead to strategic execution; and conceptual endorsement did not guarantee operational behaviour.

These findings suggest that the challenge is not teaching students that AI should not replace thinking, as they already know this, but providing them with structured mechanisms to regulate how AI is used within learning processes. In response, the study introduces the TACO framework (Think–Ask–Check–Own), a process-oriented model designed to operationalise the boundary between cognitive support and cognitive substitution. By shifting attention from ethical awareness to cognitive regulation, the study contributes a learner-grounded approach to sustaining AI as a dynamic cognitive partner in education.

## Introduction

### AI as Mediated Cognition: Theoretical Foundations and Emerging Tensions

Generative AI (GenAI) has rapidly become an everyday learning companion for many students, yet research on how students regulate AI usage are limited. Scholarships on AI in education have consistently reported the dual framing that learners and educators can articulate both benefits (efficiency, accessibility, support) with usages such as brainstorming, drafting, summarising, and explaining concepts, while simultaneously warning that AI can mislead with hallucinations, deskill with overreliance and integrity concerns (Alwaqdani, 2025; Chan & Hu, 2023; Chan & Lee, 2023; Sardi et al., 2025). Yet, what remains under-specified is the interactional regulation problem: learners may "know the rule" (e.g., don't let AI replace thinking; it is unethical to copy the AI output) but still struggle to operationalise that rule in concrete, repeatable learning routines.

Classic learning theories already provide the language for why AI pedagogy matters. Sociocultural accounts argue that learning is mediated by tools and signs, and that tools reshape what learners can do and how they do it (Gibson et al., 2023; Vygotsky, 1978) Distributed cognition extends this by treating cognition as spread across people, artefacts, and environments rather than contained solely "in the head" (Hutchins, 1995; Sidorkin, 2025). Activity theory emphasises that tool use is goal-directed and embedded in activity systems shaped by rules, community, and division of labour (Engeström, 1987; Li et al., 2025). Within these traditions, the educational significance of GenAI is not just that it provides answers, but that it can become part of the learner's cognitive system, restructuring planning, sense-making, and monitoring (Chan, 2026; Xu, 2025). Self-regulated learning (SRL) theory further clarifies that effective learning depends on cycles of forethought (goals and plans), performance (strategies and monitoring), and self-reflection (evaluation and adaptation) (Chiu, 2024; Zimmerman, 2000). GenAI can plausibly support SRL by reducing friction in planning, providing feedback, or modelling strategies but it can also undermine SRL if it becomes a shortcut that bypasses goal setting, monitoring, and reflection (Chan, 2026; Pergantis et al., 2025). Cognitive load theory helps explain why this substitution drift can be tempting: learners are sensitive to effort costs, and tools that reduce extraneous cognitive load may improve learning, while tools that remove germane processing can degrade learning (Meng et al., 2025; Sweller, 1994). Together, these theories suggest a central educational question: not whether GenAI is used, but whether its use supports or displaces learners' regulation of thinking.

A systematic review focused on assessment highlights the tension between legitimate support (feedback, drafting help) and academic integrity threats (outsourcing assessed work), and notes that institutions have struggled to define acceptable boundaries (Lee et al., 2025). Policy-facing organisations echo these themes. UNESCO's guidance on GenAI in education and research frames GenAI as potentially beneficial for learning and teaching but stresses the need for governance, transparency, and learner protection, especially around accuracy, privacy, bias, and responsible use norms (UNESCO, 2023). OECD's Digital Education Outlook 2023 similarly warns that GenAI "could also lead to the atrophy of human skills and agency and an increased dependency on the availability of AI and other technology, including for skills that are essential for success and well-being" (OECD, 2023). These macro-level reports reinforce a key point for classroom research: awareness campaigns (such as "AI can be wrong"; "don't rely too much") are necessary but are insufficient, because the core challenge is how learners manage the boundary between assistance and outsourcing during real tasks.

**The Awareness–Regulation Gap in AI-Supported Learning**

Across the literature, many studies document that students can verbalise risks (hallucinations, bias, deskilling, plagiarism), yet still report using GenAI for tasks that could plausibly substitute for effortful learning (e.g., drafting whole paragraphs, generating notes, producing structures for essays) (Chan, 2025; Sardi et al., 2025). Reviews often describe this as "overreliance," but overreliance is frequently treated as a trait (students rely too much) or an ethical issue (students cheat or AI-giarism) (Chan, 2025; Nguyen & Goto, 2024) rather than a cognitive-regulatory phenomenon (students lack stable procedures for deciding when, how, and how much to rely) (Zhai et al., 2024; Zhang et al., 2024). This is where my "diet" analogy becomes analytically useful: people can endorse healthy eating beliefs ("I should not eat cake every day") while still failing to enact behavioural control when under stress, time pressure, or temptation. Similarly, students may endorse the belief "I should not use AI to replace thinking," but under deadline pressure they may default to the most immediately efficient use pattern: letting AI generate the structure, phrasing, or even full answer. Importantly, this does not necessarily mean students are unaware or unethical; it may simply indicate that they lack a structured interaction routine that stabilises regulation at the moment of use (Abbas et al., 2024; Acosta-Enriquez et al., 2025). Also, learners can feel like they "know" because the output looks polished – a Dunning-Kruger Effect (Fernandes et al., 2024), even if they did not engage in the cognitive work that would produce durable learning. In high-pressure educational environments, where students face heavy workloads and limited time, efficiency-driven AI use may be particularly appealing, increasing the risk of superficial learning.

Scholars warn that GenAI may lead students to bypass critical thinking and rely on AI-generated outputs without sufficient evaluation (Chan & Colloton, 2024; Melisa et al., 2025; Wu, 2024). If verifying AI output requires additional time and expertise, learners may also skip it (Kim et al., 2025; Martín-Moncunill & Alonso, 2025), particularly when they trust the system or when they feel the output appears confident. The speed and fluency of AI responses may create an illusion of understanding, a phenomenon related to the "fluency heuristic," where easily processed information is perceived as more accurate (Alter & Oppenheimer, 2009).

One might argue that SRL theory already provides what is needed: teach goal setting, monitoring, and reflection. But GenAI introduces a specific interaction problem that general SRL instruction may not directly address: learners must make repeated, micro-level decisions about delegation (what to hand off), verification (what to check), and ownership (what they can legitimately claim as "their" knowledge) (Xu, 2025). Classic mediation and distributed cognition theories explain why tools matter, and cognitive load theory explains why substitution is tempting, but none of these frameworks specify a practical interaction model for regulating AI-as-partner in everyday learning moves. UNESCO similarly provides high-level guidance and principles, but principles alone do not guarantee behavioural routines (UNESCO, 2023). What we understand from research is not enough to demonstrate that students recognise risks, and it is not enough to warn them not to outsource thinking. The missing piece is a structured, teachable interaction model that translates conceptual awareness into procedural regulation.

These concerns underscore the importance of metacognition, defined as awareness and regulation of one's cognitive processes (Flavell, 1979). Metacognitive engagement ensures that learners do not simply accept AI outputs but reflect on their validity and integrate them into their own understanding. Educational frameworks that emphasise "human-in-the-loop" AI usage highlight the need for learners to remain epistemic agents who judge correctness and meaning (Memarian & Doleck, 2024). This aligns with calls for AI literacy education to include not only technical knowledge but also critical and ethical dimensions of AI use (UNESCO, 2023).

Chan (2026) identified nine learner-described uses of AI and demonstrated that it is not the technology itself that determines whether AI functions as cognitive extension or cognitive substitution. According to the study (Chan, 2026), AI is positioned along a spectrum between two contrasting learner positionings: as a cognitive partner that supports and extends thinking, and as a substitute that performs cognitive work on the learner's behalf. Students explicitly identified this support–substitution boundary, recognising that the same AI function can either scaffold sense-making or displace effortful processing, depending on how learners position the system within their cognitive activity.

The present study builds on this insight by examining a related but distinct question: **How do students regulate the boundary between using AI as support and relying on it too much in their actual learning practices?**

Where Chan (2026) focused on how learners conceptualise AI's role, this study focuses on how learners operationalise that role. Specifically, it investigates whether students demonstrate structured forms of cognitive regulation when interacting with AI as a learning partner. This focus is necessary because conceptual awareness of the support–substitution boundary does not automatically indicate that students regulate it consistently in practice.

The findings of this analysis provide the empirical basis for introducing the TACO framework (Think–Ask–Check–Own), a process-oriented model designed to articulate and measure structured human–AI cognitive regulation. Rather than assuming that ethical awareness ensures responsible AI use, the framework identifies specific stages of regulated interaction that help learners maintain AI as cognitive support rather than cognitive replacement.

**Methodology**

**Research Design**

This study adopts a qualitative interpretive design to examine how learners articulate awareness, boundaries, and regulatory strategies in relation to AI use. The analysis is theory-informed but inductive in execution, allowing regulatory patterns to emerge organically from the data. The aim was not to evaluate whether students used AI "correctly," but to examine how they themselves conceptualised and operationalised the boundary between AI assistance and cognitive substitution.

**Data Collection**

Data were collected from 133 secondary students enrolled in an AI literacy programme conducted in a Hong Kong educational context. As part of the programme, students were invited to submit written piece (400 - 800 words) responding to prompts about:

- Their personal experiences how they use AI for learning,
- Perceived benefits and risks of AI in education,
- How AI may change the roles of teachers and students,
- Challenges and ethical considerations.

Participation was voluntary and responses were collected for research purposes with appropriate parent and institutional consent procedures. All data were anonymised prior to analysis.

Importantly, students were not explicitly asked to describe regulatory strategies. The prompts focused broadly on experience, benefits, and concerns. This methodological feature is significant: descriptions of regulation, or the absence of regulation, emerged naturally rather than being elicited through direct questioning. The dataset therefore captures students' spontaneous framing of AI use.

**Analytic Strategy**

Data analysis followed an iterative qualitative interpretive process (Srivastava & Hopwood, 2009) grounded in close reading and constant comparison. The first phase involved repeated reading of all 133 responses to gain familiarity with the data. Open coding was conducted inductively, focusing on segments in which students described (a) perceived risks of AI in learning, (b) their actual AI-use practices, and (c) statements about boundaries, independence, or self-control. Codes were descriptive rather than evaluative and remained close to participants' wording. Recurring categories included references to overreliance, loss of independent thinking, misinformation, bias, privacy concerns, academic dishonesty, efficiency, speed, stress reduction, idea generation, summarisation, feedback delegation, and planning outsourcing.

During the second phase, constant comparative analysis (Glaser, 1965) was used to examine patterns across the findings. Codes were compared within and across cases to identify recurring tendencies rather than isolated examples. Two researchers assisted and examined the codes to ensure the patterns identified were correct. Through this process, four analytically distinct clusters became visible.

- First, many students demonstrated awareness of potential risks associated with AI use, including cognitive dependency, over-reliance, superficial learning, misinformation, and ethical concerns.
- Second, students frequently expressed intentions to use AI responsibly, emphasising independence, balance, and self-control in their learning practices.
- Third, emotional language highlighting AI's efficiency, speed, and ability to reduce stress appeared consistently across responses.
- Fourth, students' behavioral descriptions of actual AI use often involved outsourcing cognitive tasks, such as summarising, organising ideas, evaluating information, or generating solutions, while providing limited evidence of verification processes or deliberate self-regulation strategies.

Following identification of these four cross-case patterns, a secondary analytic step was conducted focusing specifically on the behavioral delegation excerpts. These excerpts were re-examined through a process lens to determine at which stage of learning AI was being introduced and what regulatory mechanisms, if any, were described. This process-oriented reanalysis led to categorisation of behavioral patterns into four phases:

1. Think (T) addresses insufficient pre-engagement cognition.
2. Ask (A) addresses unstructured delegation patterns.
3. Check (C) addresses limited verification procedures.
4. Own (O) addresses weak articulation of internalisation and transfer.

Crucially, the TACO framework was not used to code the entire dataset deductively. The framework was created to explain the gap we observed between students' awareness of AI risks and how they actually used AI in practice. the main regulatory gaps found in students' descriptions of their behaviour. These gaps included: limited evidence of independent thinking before using AI, overly broad or unrestricted delegation to AI, weak or unclear checking of AI outputs, and little reflection on whether learning was truly understood and internalised.

In other words, the analysis moved in two steps. First, we identified and described patterns in the data, such as students' awareness of risks, their strong statements about independence, their positive feelings about AI's efficiency, and their tendency to delegate tasks. Second, we used these patterns to build an explanatory model, the TACO framework to help make sense of the gap between what students say and what they actually do. This approach keeps the findings closely connected to the real data, while also providing a clear theoretical structure to explain the regulatory gap seen across students.

**Findings**

**Students Perception of Cognitive Awareness and Regulation**

Analysis of the 133 student reflections revealed a consistent four-part pattern: (1) strong conceptual risk awareness, (2) explicit rhetorical endorsement of independence, (3) emotionally positive framing of AI efficiency, and (4) limited articulation of structured cognitive regulation in practice. Together, these patterns illuminate a mismatch between what students say about AI and how they describe using it.

1. **Conceptual Risk Awareness: Clear Recognition of Substitution and Developmental Risks**

Students demonstrated sophisticated awareness of AI-related risks. Many explicitly articulated the danger of cognitive substitution. For example, one student noted that "*students can just copy what the AI tool provided them, without actually thinking and questioning about the possible answers*," directly identifying the risk of cognitive bypass. Others warned that overreliance could "*result in the neglect of important traditional teaching methods and the development of critical thinking and problem-solving skills*," reflecting awareness of developmental consequences.

Epistemic unreliability was frequently acknowledged. Students stated that "*AI may not be 100% accurate*," that it "*can sometimes provide misinformation*," and that "*the content generated may not always be accurate and not always trustworthy*." Several referenced contextual misalignment, observing that AI suggestions "*may not align perfectly with the specific structure required for an exam.*" These statements indicate awareness of hallucination risk and contextual limitations.

Concerns extended beyond accuracy to broader developmental and ethical implications. Students warned that excessive use "*may lose the ability to think independently*," "*weaken critical thinking, creativity and even social skills*," and "*inhibit their learning or development.*" One reflection emphasised that "*finding the correct answer is not the reason of education*," distinguishing between product and process. Ethical awareness also appeared in statements about "*privacy concerns and data security issues*" and the risk that "*bias in AI can cause some groups to be treated unfairly.*"

Collectively, these statements demonstrate that students understand the conceptual boundary between AI assistance and AI substitution. The support–substitution distinction is cognitively present in their accounts.

## 2. Stated Belief in Independent Learning: Expressing the Importance of Human Agency

Beyond risk awareness, students strongly endorsed human agency in normative terms. Many asserted that "*AI should serve as an aid rather than a substitute*," that "*AI should never replace teachers*," and that "*independent learning and teaching… should never be replaced*." These statements reflect clear conceptual positioning of AI as supportive rather than central.

Students frequently invoked moderation language. Phrases such as "*we just have to strike a balance*," "*use AI wisely*," and "*it is essential to strike a balance*" appeared repeatedly. Moral framing was also common: "*people have to be self-controlled*," and "*use AI properly instead of misusing it to do everything for them*." One student emphasised ownership, stating, "*I hardly make the AI do the work of building complete essays for me, since I also want the works to be mine and mostly based on my effort*."

Importantly, these rhetorical positions reinforce the support–substitution boundary at a normative level. However, while they clearly express values, they rarely describe structured procedures for maintaining those boundaries in practice. The language is principled but abstract.

## 3. Emotional Efficiency: AI as Relief, Speed, and Companion

In contrast to cautionary awareness, students frequently described AI in emotionally positive and efficiency-driven terms. AI was framed as providing cognitive relief: "*In a blink of an eye… without the tormenting process of thinking*." Speed and immediacy were repeatedly emphasised: "*AI generated it in seconds*," "*it gives feedback in seconds*," and "*there's little to none waiting time for getting answers*."

Efficiency was positioned as academic optimisation. Students described importing entire textbooks to "*summarise the whole book in a mere minute*," using AI to "*streamline my study process*," and receiving "*instant feedback*" to "*learn from them quickly*." AI was framed as stress reduction: "*It greatly improves my work efficiency and lessens my stress drastically*."

Anthropomorphic and relational language also emerged. AI was described as "*a good friend throughout my learning*," "*my knight in shining armor*," and even "*a permanent friend who is always here for me whenever I need help*." One student stated, "*It saves my life sometimes*." These expressions suggest emotional reliance and attachment beyond purely instrumental use.

This efficiency-oriented framing introduces tension. While students conceptually reject substitution, they simultaneously valorise speed, automation, and relief from cognitive effort, conditions that make substitution more likely.

## 4. Behavioral Gap: Delegation Without Structured Regulation

The most analytically significant pattern emerges in students' descriptions of actual AI use. Across the responses, AI was frequently used for initiation, planning, synthesis, evaluation, and solution generation, yet without detailed articulation of regulatory strategies.

Students described asking ChatGPT "*for some idea,*" uploading drafts and "*allowing it to evaluate,*" requesting "*long and detailed explanations*," and seeking feedback after drafting essays. Others reported outsourcing planning: "*I could ask Poe to help me create a timetable… and even ask it to

*arrange the study sessions.*" Synthesis was delegated through statements such as "*I uploaded my learning resources to AI, which then created customised study notes*" and "*I use AI to summarise contents or simplify paragraphs.*"

Direct solution access was normalised: "*By simply taking a picture of a math equation… it will show the detailed solutions.*" Students acknowledged that "*AI gives students a direct answer to their struggles,*" yet rarely described attempting solutions independently first.

Even when substitution dynamics were recognised, procedural safeguards were absent. Statements such as "*students can just use AI to generate essays*" or "*I can just ask ChatGPT*" were not accompanied by articulated boundaries, verification routines, or internalisation strategies.

Importantly, these four clusters did not operate independently. Taken together, the findings reveal a consistent misalignment. Students understand AI's risks conceptually. They defend independence rhetorically. They appreciate AI's efficiency emotionally. But they rarely describe structured cognitive regulation behaviorally. Explicit descriptions of repeatable strategies, such as attempting tasks independently before consulting AI, defining criteria for evaluating AI output, or articulating how AI-assisted learning was internalised were comparatively uncommon.

Awareness does not automatically translate into procedural regulation. The boundary between support and substitution is acknowledged, yet not systematically operationalised. This empirical pattern forms the basis for the framework introduced in the next section.

**Discussions**

The findings of this study reveal a consistent and theoretically significant mismatch between students' conceptual understanding of AI-related risks and their ability to regulate AI use in practice. Across 133 written responses, students demonstrated clear awareness that GenAI can mislead, reduce independent thinking, and undermine learning when overused. They also articulated strong normative commitments to human agency, frequently asserting that AI should "*serve as an aid rather than a substitute*" and that "*finding the correct answer is not the reason of education.*" At the same time, students described AI in emotionally positive terms, emphasising its speed, convenience, stress reduction, and constant availability. Yet, when reflecting on how they actually used AI during learning tasks, students rarely articulated explicit, repeatable strategies for managing the boundary between assistance and cognitive outsourcing.

This awareness–regulation gap suggests that the central challenge in AI-supported learning is not a lack of ethical belief or conceptual understanding. Rather, it lies in the absence of structured cognitive regulation during moment-to-moment interaction with AI systems. Students appear to "know the rule" that AI should not replace thinking, but struggle to operationalise that rule under conditions of time pressure, cognitive load, and efficiency-driven decision-making. This pattern aligns with prior research showing that learners can articulate risks such as hallucination, overreliance, and deskilling while still engaging in AI use that plausibly substitutes for effortful learning (Chan & Hu, 2023; Sardi et al., 2025).

Importantly, the findings indicate that overreliance should not be interpreted solely as an ethical failure or a motivational deficit. Instead, it can be understood as a cognitive-regulatory problem.

From a self-regulated learning perspective, effective learning depends on coordinated cycles of forethought, strategic action, monitoring, and reflection (Zimmerman, 2000). The present data suggest that while students may endorse these principles at a conceptual level, they often lack concrete procedures for enacting them when AI is introduced as a powerful and efficient cognitive resource. Similarly, distributed cognition and sociocultural theories explain why AI can become part of learners' cognitive systems, but they do not specify how learners should actively regulate this partnership to prevent cognitive substitution (Hutchins, 1995; Vygotsky, 1978).

These findings extend prior work that positions AI along a spectrum between cognitive support and cognitive substitution (Chan, 2026). While earlier research demonstrated that students can recognise this boundary conceptually, the present study shows that recognising the boundary does not ensure that it is consistently regulated in practice. The gap identified here is therefore not epistemic but procedural: students lack a stable interaction model that guides how, when, and to what extent AI should participate in different stages of learning.

Taken together, the results point to a need for frameworks that move beyond awareness-raising and ethical injunctions toward structured regulation of human–AI interaction. In the sections that follow, we introduce the TACO framework (Think–Ask–Check–Own) as a process-oriented model designed to operationalise cognitive regulation during AI-supported learning. Rather than prescribing whether AI should or should not be used, the framework addresses how learners can remain epistemic agents while engaging AI as a cognitive partner rather than a cognitive replacement.

**Students' Behavioral Patterns and the Development of TACO**

The behavioral patterns presented above reveal a consistent absence of structured cognitive regulation in students' descriptions of AI use. While students frequently articulated awareness of AI's risks and expressed normative commitments to independence, their procedural accounts rarely specified how they regulate the boundary between assistance and substitution in practice. Across the responses, AI engagement was often initiated immediately "*I can just ask ChatGPT*", tasks were externally structured "*I use AI to generate ideas and structure my essays*", evaluative judgment was delegated without articulated criteria "*Uploading my work… allowing it to evaluate*", and outputs were accepted rapidly "*It gives feedback in seconds*" without described verification procedures. Similarly, large-scale summarisation and solution generation were commonly reported, yet few students detailed processes of reconstruction, transfer, or self-testing that would indicate internalisation of understanding.

Importantly, the issue is not that students lack awareness. As demonstrated earlier, many explicitly cautioned against overreliance and endorsed the principle that "*AI should serve as an aid rather than a substitute*." However, these positions were primarily declarative rather than procedural. Statements such as "*we just have to strike a balance*" or "*use AI wisely*" reflect ethical orientation but do not constitute structured regulatory strategies. The data therefore suggest a recurring awareness–regulation gap: students recognise the conceptual boundary between cognitive support and cognitive replacement, yet rarely articulate repeatable mechanisms for managing that boundary during actual AI interaction.

This empirical pattern informed the development of the TACO framework. Rather than being imposed a priori, TACO was constructed deductively from the regulatory absences observed across the corpus. Each component corresponds directly to a recurrent gap identified in the findings. The absence of articulated independent cognitive initiation informed **Think**, a structured pre-engagement phase requiring learners to define goals and attempt initial reasoning before AI input. The pattern of unconstrained delegation informed **Ask**, emphasising bounded, intentional prompting rather than wholesale outsourcing. The lack of verification procedures informed **Check**, introducing explicit validation and reconstruction steps. Finally, the limited evidence of internalisation or accountability informed **Own**, foregrounding cognitive ownership, transfer, and reflective consolidation.

TACO is not an ethical statement about responsible AI use, nor is it a prohibition-based model designed to restrict technology. TACO operationalises the support–substitution boundary that students already recognise conceptually but do not consistently regulate behaviorally. The framework therefore responds not to a deficit of ethical awareness, but to the absence of structured interaction models guiding how learners engage AI as a cognitive partner. The following section introduces TACO as a procedural model designed to transform abstract caution into actionable cognitive regulation.

**The TACO Framework: A Procedural Model for Regulating AI as Cognitive Partner**

TACO integrates insights from multiple theoretical traditions while addressing a practical regulatory need. From sociocultural theory (Vygotsky, 1978), AI may function as a mediational tool within the learner's activity system. However, mediation enhances development only when externally supported processes are internalised rather than permanently externalised. From distributed cognition (Hutchins, 1995), cognitive processes can extend across human and technological systems. Yet distribution does not eliminate responsibility; it shifts the need toward coordination and control. From self-regulated learning theory (Zimmerman, 2000), effective learning requires planning, monitoring, and evaluation. AI integration intensifies this demand, as learners must now regulate not only their cognition but also their delegation. From a cognitive load perspective (Sweller, 1994), AI can reduce extraneous load. However, premature reduction of germane load risks undermining schema construction and long-term transfer. Despite these theoretical resources, no existing framework specifies how students should regulate AI interaction procedurally across stages of task engagement. TACO responds to this absence by proposing a four-phase regulatory cycle embedded directly within AI use.

**TACO as a Four-Phase Interaction Cycle**

TACO conceptualises AI engagement as sequential yet recursive. The phases are not rigid steps but structured checkpoints that prevent unintentional substitution drift. To make visible the regulatory absences identified across cases, Table 1 presents illustrative excerpts aligned with the four emergent phases that later informed the TACO framework. The table does not represent exhaustive coding but highlights recurring patterns in which cognitive delegation occurred without corresponding procedural safeguards. Table 2 and figure 1 present the TACO framework as a four-phase regulatory model, outlining the purpose, learner actions, and substitution risks addressed at each stage (Think, Ask, Check, Own) to operationalise structured human–AI cognitive partnership.

| TACO Phase | Illustrative Verbatim Excerpt | Observed Behavioral Pattern | Regulatory Gap Identified |
|---|---|---|---|
| **T – Think (Pre-engagement cognition)** | "I can just ask ChatGPT." | Immediate AI initiation | No independent cognitive attempt described |
| | "I use AI to generate ideas and structure my essays." | Task initiation externally delegated | No evidence of self-generated outline or prior reasoning |
| | "By simply taking a picture of a math equation… it will show the detailed solutions." | Instant solution access | No productive struggle or problem framing articulated |
| **A – Ask (Structured delegation)** | "Give me long and detailed explanations…" | Broad, unconstrained request | No boundary conditions specified |
| | "Uploading my work… allowing it to evaluate…" | Evaluative judgment delegated | No evaluation criteria articulated |
| | "I use AI to automate this task." | Automation prioritized | No articulation of what cognitive work should remain human |
| **C – Check (Verification & reconstruction)** | "It gives feedback in seconds." | Speed equated with learning | No verification process described |
| | "AI may not always be correct." | Conceptual awareness of inaccuracy | No operational checking procedure described |
| | "Students should always check the validity of sources." | Normative endorsement of verification | Declarative claim without procedural detail |
| **O – Own (Internalisation & accountability)** | "I use AI to summarize contents or simplify paragraphs." | Compression replaces elaboration | No reconstruction in own words described |
| | "It helped me to generate the layout of a particularly difficult essay in less than 5 seconds." | Structural generation delegated | No rewriting or consolidation process articulated |
| | "It saves my life sometimes." | Emotional reliance | No reflection on independent competence or transfer |

Table 1. Behavioral Patterns Underlying the TACO Framework

| Phase | Core Purpose | What the Learner Does | What It Prevents |
|---|---|---|---|
| **T — THINK** | Human-first cognitive initiation | • Clarifies the task in own words<br>• Activates prior knowledge<br>• Attempts an initial idea or draft<br>• Identifies knowns and unknowns | Premature outsourcing;<br><br>starting with AI instead of thinking |
| **A — ASK** | Bounded and intentional delegation | • Asks for hints, explanations, or feedback<br>• Sets clear limits in prompts<br>• Requests guidance rather than full solutions<br>• Specifies criteria or constraints | Unrestricted delegation;<br><br>AI doing the entire task |
| **C — CHECK** | Verification and epistemic control | • Cross-checks facts with other sources<br>• Tests logic or reasoning steps<br>• Identifies inconsistencies or bias<br>• Reconstructs answers in own words | Blind acceptance;<br><br>fluency bias;<br><br>misinformation |
| **O — OWN** | Internalisation and accountability | • Writes final response in own voice<br>• Explains the concept without AI<br>• Applies learning to a new task<br>• Reflects on what was personally learned | Performance without understanding;<br><br>externalised cognition |

Table 2: The TACO Framework for Regulating AI as a Cognitive Partner

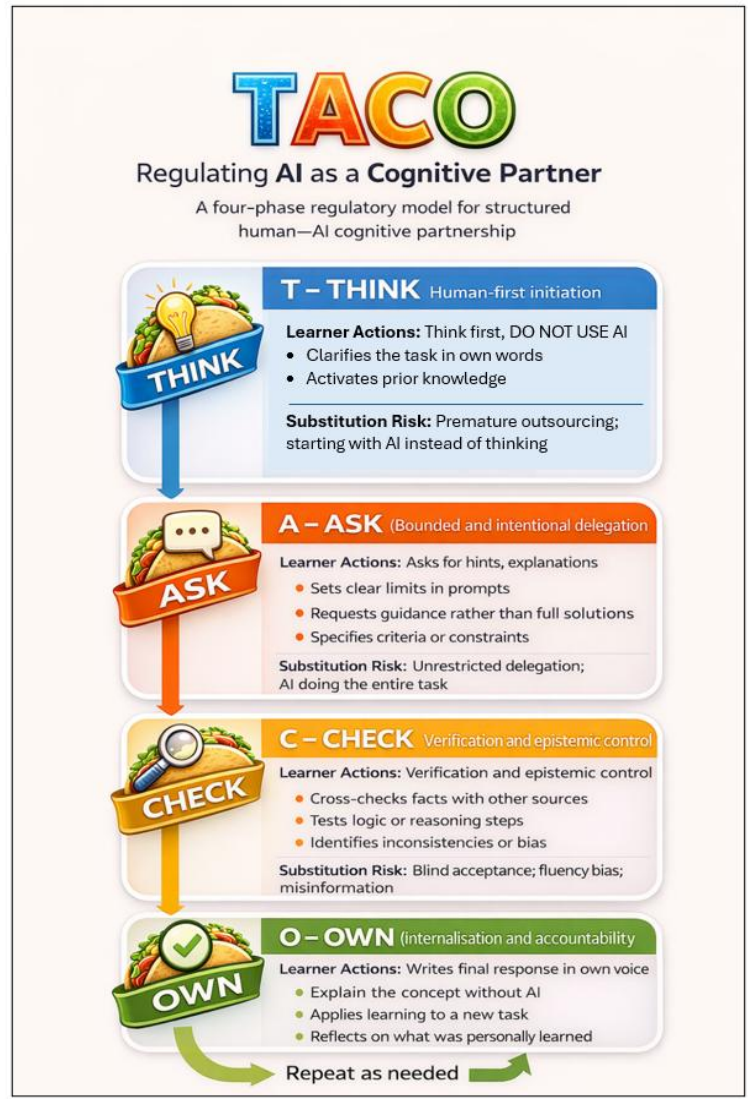

Figure 1: The TACO Framework for Regulating AI as a Cognitive Partner

1. **THINK: Human-First Cognitive Initiation**

The Think phase requires learners to initiate cognitive engagement before consulting AI. This includes clarifying task goals, activating prior knowledge, generating preliminary ideas, and identifying specific areas of confusion.

From the findings, students described immediate prompting or delegation without evidence of independent attempt. Such patterns suggest weakened cognitive initiation. When AI becomes the starting point rather than a scaffold, the learner's role shifts from agent to recipient.

The Think phase puts the learner's thinking first. It makes sure the AI supports an idea that's already forming, instead of doing the thinking for the learner. This approach matches self-regulated learning research, which shows that students learn better when they plan ahead and analyse the task before starting (Zimmerman, 2000). It also supports deeper learning by encouraging students to begin building their own understanding before getting help (Sweller, 1994). Operationally, the Think dimension involves:

- Articulating the task in one's own words
- Attempting a partial solution
- Listing known and unknown elements
- Generating at least one independent idea

This phase prevents premature outsourcing and establishes the learner as cognitive initiator.

**2. ASK: Structured Delegation Rather Than Substitution**

The Ask phase regulates how AI is consulted. The empirical findings show frequent use of AI for idea generation, summarisation, and evaluation, yet rarely with described boundaries or constraints. Delegation often appears open-ended "*I ask AI to generate…*", increasing substitution risk.

Ask introduces bounded prompting. Rather than requesting full solutions, learners are guided to request:

- Hints instead of answers
- Explanations instead of completions
- Feedback on drafts rather than generated replacements
- Alternative perspectives rather than definitive conclusions

This phase operationalises the support–substitution boundary. AI is positioned as scaffold rather than solver.

The Ask phase is based on the idea that thinking can be shared between people and tools (Hutchins, 1995). But it's important to be clear about who is responsible for what. The learner should stay in charge, even when getting help. It also connects to the idea of scaffolding (Vygotsky, 1978), where support helps someone learn, but doesn't replace their own thinking.

Without clear rules for how to "Ask," AI can easily move from helping to doing the work itself. The Ask phase makes sure the AI supports the task without taking it over.

**3. CHECK: Verification and Epistemic Control**

Across the dataset, conceptual awareness of AI inaccuracy and hallucination was common. However, explicit verification procedures were inconsistently described. Students frequently stated that AI "*may not always be correct*" without detailing how they evaluate outputs.

The Check phase embeds verification as a mandatory regulatory checkpoint. It requires learners to:

- Compare AI output with prior understanding
- Cross-check information with trusted sources
- Identify inconsistencies or overgeneralisations
- Reconstruct explanations in their own words

The Check phase gives control back to the learner. It helps prevent fluency bias, the tendency to believe something is correct just because it sounds clear or confident (Alter & Oppenheimer, 2009). It also supports self-regulated learning (Zimmerman, 2000), where learners actively monitor and evaluate their own understanding. From a learning perspective, Check keeps students from passively accepting information. Instead of just absorbing what the AI says, they must question it, test it, and connect it to what they already know, developing their evaluative judgement and critical thinking.

Check isn't about being skeptical of everything. It is about carefully evaluating the information so the AI stays a helpful tool, human leads with judgement within the human-AI partnership.

### 4. OWN: Internalisation, Transfer and Accountability

The Own phase addresses the most subtle regulatory gap identified in the findings: internalisation. Students often described using AI effectively yet rarely articulated how knowledge was consolidated independently of AI.

Own ensures that learning remains located within the student. It requires reflection on:

- What was learned
- What AI contributed versus what the learner constructed
- Whether the learner could complete a similar task independently
- How understanding transfers to new contexts

This phase means that thinking supported by others (or by AI) should eventually become the learner's own thinking (Vygotsky, 1978). The goal is to ensure the knowledge is truly understood and can be used independently. It also strengthens the learner's sense of responsibility and agency.

Without Own, AI may enhance performance but not competence. Tasks may be completed successfully, yet cognitive ownership remains externalised.

### TACO as Boundary Regulation

The central contribution of TACO is not its individual components but its function as boundary regulation. It structures the dynamic tension between cognitive extension and cognitive substitution.

- Think prevents premature externalisation.
- Ask regulates the scope of delegation.
- Check reasserts epistemic control.
- Own secures internalisation and transfer.

Together, these phases transform a moral injunction ("*Do not rely too much on AI*") into a cognitive procedure. They provide learners with an actionable framework for managing AI participation across task stages.

This directly addresses the empirical finding that:

> Awareness ≠ Regulation
> Ethical belief ≠ Strategic execution
> Conceptual endorsement ≠ Operational behavior

Students do not require further warnings about AI risks. They require structured interactional design.

**Implications for Human–AI Partnership**

TACO reframes AI literacy from knowledge about AI to regulation of AI. It positions learners not as passive users nor as technology skeptics, but as cognitive orchestrators within distributed systems.

The framework suggests that effective human–AI partnership depends not on limiting AI access, but on embedding metacognitive checkpoints within interaction cycles. In doing so, it extends theoretical models of mediation and distributed cognition into practical educational design.

Rather than asking whether AI should be used, TACO asks:

- How is AI entering the task?
- Who initiated cognition?
- Who verified outputs?
- Where is understanding located?

By structuring these questions into a procedural cycle, TACO makes visible the regulatory moves that were largely absent in students' naturalistic accounts.

**Conclusion**

This study introduced the TACO framework for assessing cognitive regulation in AI-supported learning. Grounded in qualitative evidence, the instrument operationalises the boundary between cognitive support and cognitive substitution by translating abstract principles into structured interaction stages: Think, Ask, Check, and Own.

The empirical findings revealed a consistent and analytically important pattern across the corpus.

- Students understand AI's risks conceptually.
- They defend independence rhetorically.
- They appreciate AI's efficiency emotionally.
- But they rarely describe structured cognitive regulation behaviorally.

Students explicitly acknowledged that AI can "*provide misinformation*," "*hinder critical thinking*," and lead to "*overreliance*." They articulated that AI should "*serve as an aid rather than a substitute*" and that "*finding the correct answer is not the reason of education*." At the same time, they framed AI as efficient, immediate, stress-reducing, and even emotionally supportive, describing it as a "*good friend*" or something that "*saves my life sometimes*." However, when describing actual AI use, students rarely articulated repeatable procedures for initiating independent cognition, constraining delegation, verifying outputs, or internalising understanding.

This awareness–regulation gap forms the central contribution of the study. The problem is not that students lack ethical belief or conceptual understanding. They already know that AI should not replace thinking. What is missing is a structured mechanism for regulating the boundary between assistance and outsourcing in practice.

The TACO framework responds directly to this empirical gap. By sequencing AI interaction into four regulatory phases, Think (human-first cognition), Ask (bounded scaffolding), Check (verification), and Own (internalisation), the model transforms abstract caution into actionable cognitive regulation. In doing so, TACO reframes AI integration in education. The issue is not whether students use AI, but how they regulate it.

Several limitations must be acknowledged. First, the instrument relies on self-report. Although items were derived deductively from qualitative patterns, responses may be influenced by social desirability or inaccurate metacognitive judgment. Future research should triangulate TACO scores with behavioral trace data, such as prompt logs or revision histories. Second, the qualitative corpus was derived from a specific educational context. Cultural, disciplinary, and developmental differences may influence AI interaction patterns. Cross-context validation is necessary to examine generalisability. Finally, while TACO is modeled as a staged process, real-world AI interaction may be recursive or nonlinear. Further refinement may explore dynamic interaction modeling.

Theoretically, TACO contributes to the literature on human–AI collaboration by shifting focus from prohibition to regulation. Existing discourse often emphasises preventing misuse. The present findings demonstrate that students already recognise misuse risks. The missing component is procedural regulation. Pedagogically, TACO provides a diagnostic framework. Educators can identify distinct patterns of AI engagement from scaffolded partnership to cognitive outsourcing and design targeted interventions. Rather than teaching students that "AI should not replace thinking," instruction can focus on how to operationalise that boundary through structured steps. At the institutional level, the framework supports proactive capacity building. Instead of relying solely on AI detection or restriction policies, institutions can cultivate regulatory competence.

Future research should focus on developing a practical instrument that helps students become aware of the support–substitution boundary *at the moment* they use generative AI. The present findings suggest that many learners can state the principle ("don't rely too much") but lack a stable mechanism for translating that principle into routine regulation. A next step, therefore, is to design process-based instrument aligned with TACO that can be used repeatedly across tasks without becoming burdensome. Finally, domain-specific adaptations of TACO may be developed for writing, mathematics, programming, and other disciplines.

The rapid integration of generative AI into education has intensified debates about overreliance and academic integrity. Yet this study demonstrates that students already grasp the fundamental risks. They conceptually recognise substitution, rhetorically defend independence, and emotionally value efficiency. What they lack is structured cognitive regulation. The future of AI-supported learning will not depend on restricting technology, but on strengthening human regulatory capacity. The TACO framework represents a step toward that goal by operationalising how learners can engage AI as a cognitive partner rather than a cognitive replacement.

**Declarations:**

Availability of data and material: The datasets used and/or analysed during the current study are available from the corresponding author on reasonable request